# Intrinsic 2D-XY ferromagnetism in a van der Waals monolayer

Amilcar Bedoya-Pinto[+1], Jing-Rong Ji[+1], Avanindra Pandeya[1],
Pierluigi Gargiani[2], Manuel Valvidares[2], Paolo Sessi[1], Florin Radu[3],
Kai Chang[1] and Stuart S.P. Parkin[1]

[1] NISE Department, Max Planck Institute of Microstructure Physics, Halle, Germany

[2] ALBA Synchrotron Light Source, Barcelona, Spain

[3] Helmholtz-Zentrum für Materialien und Energie, Berlin, Germany

Correspondence to: abedoya@mpi-halle.mpg.de, stuart.parkin@mpi-halle.mpg.de

[+] these authors contributed equally to this work

**Abstract**

Long before the recent fascination with two-dimensional materials, the critical behaviour and universality scaling of phase transitions in low-dimensional systems has been a topic of great interest. Particularly intriguing is the case of long-range magnetic order in two dimensions, once considered to be excluded in systems with continuous symmetry by the Hohenberg-Mermin-Wagner theorem. While an out-of-plane anisotropy has been shown to stabilize 2D magnetic order, this proof has remained elusive for a 2D magnet with in-plane rotational symmetry. Here, we construct a nearly ideal easy-plane system, a $CrCl_3$ monolayer grown on Graphene/6H-SiC (0001), and unambiguously demonstrate robust in-plane ferromagnetic ordering with a critical scaling behaviour characteristic of a 2D-XY system. These observations suggest the first realization of a finite-size Berezinskii-Kosterlitz-Thouless (BKT) phase transition in a large-area, quasi-freestanding, van der Waals monolayer magnet with a XY universality class; and further constitute an ideal platform to study exotic phenomena like superfluid spin transport or 2D topological in-plane spin textures- such as merons.

**One Sentence Summary:**

Discovery of the first scalable, atomically-thin van der Waals ferromagnet grown by molecular-beam epitaxy featuring an easy-plane magnetic anisotropy and a 2D-XY universality class.

**Main Text**

How physical properties change in systems with reduced dimensionalities is a scientific question that has fascinated researchers for decades. Taking advantage of recent technological advances, the possibility to design on-demand atomic-sized entities have led to a plethora of studies of truly 1 and 2-dimensional systems. In this regard, theoretical models of low-dimensional magnetism have been thoroughly revisited hand-in-hand with new experimental findings. It was early noted that the absence of long-range magnetic order in a low-dimensional system (d<3) with continuous symmetry, postulated by Mermin and Wagner *(1,2)*, could be lifted by the existence of a sizable magnetic anisotropy. A special case, thereof, are 2D systems with easy-plane anisotropy, where a special long-range order is expected to emerge due to the formation of magnetic vortex-antivortex or chiraly opposing domain-wall bound pairs, theoretically described by the Berezinskii-Kosterlitz-Thouless (BKT) formalism *(3-5)*. This implies the occurrence of a phase transition even in a 2D magnetic system with continuous rotational symmetry *O(2)*. This theory has been later adapted to account for symmetry breaking fields *(6-8)* and finite size effects *(9-10)* and spin-wave interactions with magnetic vortices *(11)*, bringing theory much closer to an experimentally realistic scenario. The seminal works of Villain *(11)*, Jose *(6)*, Szeto *(7)* and Bramwell *(9)* that followed the initial BKT theory allowed the quantification of the order parameters and critical scaling near the phase transition for each scenario. Bramwell and Holdsworth *(9,12)* showed by a renormalization group approach that the magnetization of finite-sized 2D easy-plane systems – also called 2DXY- scale with a critical exponent of $\beta= 3\pi^2/128 \approx 0.231$, defining a fingerprint of the XY universality class. Perturbations to the rotational invariance of the spins in the plane -such as the symmetry and strength of magnetocrystalline fields- have been shown to strongly influence the critical exponents and the universality class. While a uniaxial in-plane anisotropy ($XYh_2$) drives the system into an Ising-type behaviour ($\beta= 0.125$), the effect of a four-fold symmetry ($XYh_4$) is highly dependent on the crystal field strength, with critical exponents ranging between Ising and XY universality classes *(13)*. In contrast, the hexagonal symmetry ($XYh_6$) barely affects the magnetic behaviour *(4,6,7)* even in the limit of strong crystal fields *(13)*, such that the system remains in the XY class, enabling the observation of a BKT-type phase transition. A large number of experiments on quasi 2D-magnetic systems, such as monolayers on crystalline metallic surfaces or bulk layered magnets with small interplanar exchange interactions, have contributed to the assessment of the aforementioned theories (comprehensive reviews can be found in Refs. 13 and 14). However, in those cases, the substrate interaction through bonding and hybridization (monolayers on surfaces) and the -even small- amount of interplanar exchange (layered magnets) were conditions that precluded the realization of an ideal 2D system. Recent advances in the isolation and preparation of crystalline monolayers from innately layered magnets, via exfoliation *(15-17)* or molecular-beam epitaxy *(18)*, allow for more detailed investigations of low-dimensional critical phenomena. So far, ferromagnetic ordering has been demonstrated in $CrI_3$, $CrBr_3$ and $Fe_3GeTe_2$ monolayers *(15-16,18)*, all of them stabilized by a substantial out-of-plane uniaxial anisotropy that places them close to the Ising universality class, as discussed in recent reviews *(19,20)*. A careful analysis of some of these systems -such as ferromagnetic $CrBr_3$ *(21)* or antiferromagnetic $NiPS_3$ *(22)*- rather suggests a XXZ-type magnetic behaviour, the latter losing magnetic ordering in the monolayer regime. $CrCl_3$, on the other hand, is an in-plane antiferromagnet in the bulk form *(23-24)*, resulting from an alternation between the magnetic moments in each individual $CrCl_3$ layer, that are ferromagnetically aligned in-plane. Thus, arises the interesting question, whether a single-layer $CrCl_3$ would order ferromagnetically –or order magnetically at all-, considering its weak anisotropy and the lifting of interplanar exchange when reaching the monolayer limit.

Recent reports on the magnetic properties of $CrCl_3$ exfoliated flakes reach down to the bilayer regime *(25-28)*, where the antiferromagnetic interlayer exchange is still preserved, whereas the determination of the magnetic properties of a single monolayer has remained elusive. One reason are the indirect methods involved in the magnetic characterization of the monolayer flakes, which require device fabrication for tunnelling magnetoresistance *(25-28)* or Hall-micro-magnetometry experiments *(21)*. Other direct methods such as the magneto-optical Kerr-effect, suffer from small signals due to sample dimensions and unfavourable magneto-optical coefficients in $CrCl_3$ *(29)*. In this work, we circumvent these difficulties by preparing a large-area, homogeneous $CrCl_3$ monolayer on Graphene/6H-SiC(0001) by molecular-beam-epitaxy (MBE), and by measuring their magnetic properties *in-situ* via element-specific X-ray magnetic circular dichroism (XMCD). We demonstrate ferromagnetic ordering of the $CrCl_3$ monolayer with a Curie-Temperature of 13 K, and quantify the local moments both on the Cr and Cl atoms. An in-plane easy axis is clearly observed, supported by an unexpectedly large anisotropy energy; and the scaling analysis near the phase transition is consistent with a 2DXY spin system. The key factors for the realization of a nearly-ideal 2DXY magnetic system rely on a reduced (van der Waals) substrate interaction and a weak hexagonal in-plane crystal field present in the CrCl$_3$ monolayer.

The structure of the CrCl$_3$ monolayer grown on graphitized 6H-SiC(0001) by MBE features a van der Waals gap to the substrate and the formation of a crystalline layer with an in-plane hexagonal lattice, as shown in Figure 1a. The Cr atom is coordinated in an octahedral configuration to the neighbouring Cl atoms, i.e. Cr-Cl bonds are off-plane, and the Cr atoms form a honeycomb lattice. *In-situ* reflection high-energy electron diffraction (RHEED) patterns show that the CrCl$_3$ films have a surface perpendicular to the c-axis, and that the film microstructure presents in-plane twisted domains, as seen in the multiple diffraction streaks corresponding to *a* and √3*a* CrCl$_3$ lattice periodicities along a high-symmetry direction (e.g. Γ-M of Graphene, Figure 1b). This means that a strict 6-fold symmetry is no longer guaranteed on long-range length scales. Figure 1c shows a 3 µm sized scanning tunnelling microscopy (STM) topographic image of a CrCl$_3$ monolayer, indicating an homogeneous coverage of the graphene substrate. Inside a CrCl$_3$ grain, the atom resolved STM image (Fig. 1d) displays the CrCl$_3$ hexagonal lattice superimposed on a clear Moiré pattern, which corresponds to a twist angle of 23.8° (see Supplementary Fig. 1) between the CrCl$_3$ monolayer and the graphene substrate. Various Moiré patterns were observed in different grains, which further supports random in-plane grain orientations (Supplementary Fig. 2) consistent with a weak interaction with the graphene substrate. The local electronic properties of the CrCl$_3$ monolayer, mapped by scanning tunnelling spectroscopy (Figure 1d), reveal a bandgap of ~1.6 eV, which is close to what is predicted from *ab-initio* calculations *(30)* but substantially lower than the bandgap (3.0 eV) measured in bulk samples *(31)*. The Fermi-energy lies in the middle of the gap, indicating that the electronic properties of the MBE-grown monolayer are intrinsic, with a low concentration of defects/dopants and a negligible charge transfer effect from the substrate. Prior to the *in-situ* investigation of the magnetic properties of the samples, we performed X-ray absorption spectroscopy measurements to assess the chemical integrity of the CrCl$_3$ surface after transfer via an ultra-high-vacuum taxi chamber to the synchrotron beamline (see Methods for details). As can be seen in Figure 1e, we did not observe any trace of oxygen contamination but rather a well-defined Cr X-ray absorption line, with sharp multiplet peaks characteristic of a Cr$^{3+}$ ion in an octahedral configuration.

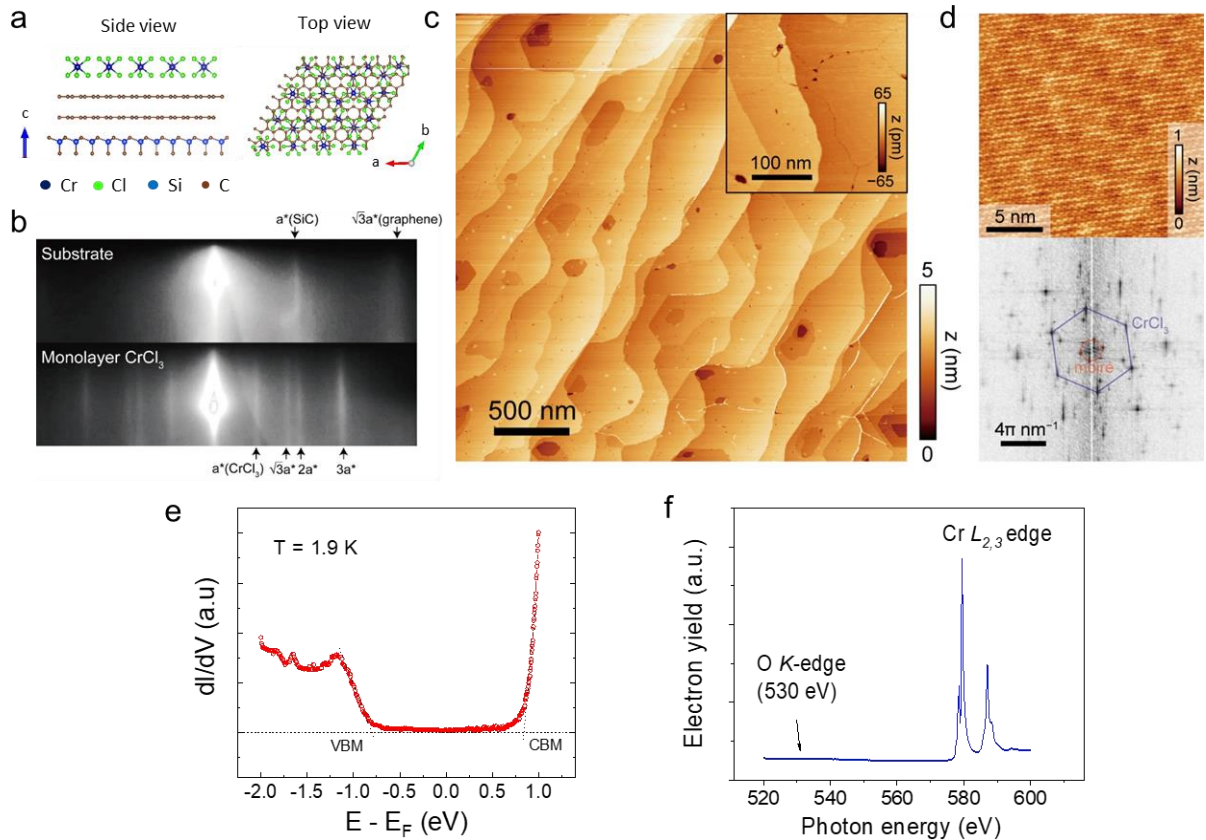

Figure 1. **Structural and electronic properties of a CrCl$_3$ monolayer.** (a) Schematic crystal structure of CrCl$_3$/Graphene/6H-SiC layers in top view and cross-section configurations. (b) In-situ RHEED pattern of the substrate and monolayer CrCl$_3$ grown by MBE, along Γ-M of Graphene (Γ-K of SiC). Streaks from different high-symmetry directions of CrCl$_3$ are observed, implying a twisted in-plane orientation of the grains. (c) STM topography of a monolayer CrCl$_3$ grown on Graphene/6H-SiC(0001), indicating an homogeneous coverage on long length scales. Setpoints: sample bias voltage V = +1.2 V, tunnelling current I = 5 pA. Inset, a magnified topography image, which reveals the grain boundaries. (d) Atom resolved image of the CrCl$_3$ lattice featuring a moiré pattern (upper panel), and its Fourier transformed image (lower panel). Setpoint, V = +0.1 V, I = 100 pA. The Moiré pattern corresponds to a 23.8° rotation between the hexagonal unit cell of CrCl$_3$ and graphene. (c) and (d) were acquired at room temperature. (e) dI/dV spectrum at the surface of a monolayer CrCl$_3$, taken at 1.9 K. The estimated bandgap is 1.6 eV obtained by linearly extrapolating the sharp increase in signal at positive and negative energies to intersect the energy axis. (f) X-ray absorption spectroscopy near the O K and Cr L$_{2,3}$ edge region, ruling out the presence of oxygen in the surface and highlighting a sharp Cr$^{3+}$ absorption white line.

Figure 2 summarizes the key magnetic features of the CrCl$_3$ monolayer as revealed from X-ray magnetic dichroism (XMCD) measurements: i) large XMCD signals at the Cr $L_{2,3}$ edge (close to 100% at $L_2$ at saturation fields) indicative of sizable magnetic moments on the $Cr^{3+}$ ions (as well as at the Cl sites, see Supp. Fig. 3), ii) a field-dependent magnetization with non-zero remanence and coercive fields typical of ferromagnetic ordering, iii) a weak but detectable magnetic anisotropy favouring an in-plane easy axis, iv) a ferro-to paramagnetic transition around 13 K. As shown in Figure 2a, the X-ray absorption spectra at the Cr $L_{2,3}$ edge taken with different photon helicities (right- and left-handed, I+ and I- for simplicity) shows a huge difference, yielding values of nearly 100% at high magnetic fields (8T) and in normal incidence geometry (field perpendicular to the plane). Interestingly, the XMCD spectra is still visible in zero-magnetic field (about 10%), which indicates the presence of a remanent magnetic moment at low temperatures (3.5 K), as depicted in Fig. 2b. This high XMCD contrast was crucial to performing measurements of the Cr $L_3$ edge at different magnetic fields (hysteresis loops) with a high level of accuracy. Fig. 2c shows the correspondence of the XMCD intensity extracted from the energy spectra (in % normalized to the XAS edge jump) to that acquired during a typical hysteresis loop measurement (for details on the measurement conditions refer to Methods). Using sum-rule analysis (Supplementary Figure 4) we extract a total magnetic moment of 2.8 µB/Cr, very close to the theoretically expected value of 3 µB for a trivalent Cr valence ($3d^3$). Although we also found a remarkably high XMCD signal (30%) at the chlorine edge (Supp. Figure 3), sum-rule analysis could not be performed due to the small energy separation of $L_3$ and $L_2$ lines resulting from the weak spin-orbit coupling of Cl. Assuming a ferromagnetic super-exchange coupling mechanism between Cr spins over Cl ligands in $CrCl_3$, it is very likely that the spin moment of Cl aligns antiparallel to that of Cr. In fact, the small reduction of the $Cr^{3+}$ magnetic moment (2.8 µB instead of 3 µB) might arise from a certain degree of p-d admixture of Cr and Cl orbitals in the super-exchange path. The resulting exchange coupling is small ($J_{ex}$ ~ 0.6 meV), in line with the low ordering temperature observed in the monolayer samples. Figure 2e displays the temperature evolution of the $Cr^{3+}$ XMCD hysteresis loops under an in-plane applied magnetic field (grazing incidence geometry, GI), showing that the squareness of the hysteresis clearly diminishes as the temperature is raised; and at 13 K, both remanence and coercive fields vanish, indicative of a ferro-to paramagnetic phase transition. In order to evaluate the magnitude of the related magnetic anisotropy, field-dependent XMCD hysteresis loops at normal and grazing incidence angles have been taken at 3.5 K (Figure 2d), evidencing a clear in-plane magnetic easy axis.

The extracted anisotropy fields, shown in the inset of Figure 2d, are on the order of 0.5 – 0.6 T (corresponding to anisotropy energies of 0.08 - 0.1 meV). These values are much larger than those anticipated from first-principle calculations of a $CrCl_3$ monolayer *(30,32-33)*, which range from 0.031 to 0.055 meV/Cr; and also appear to be slightly larger than the critical field anisotropy in the antiferromagnetic bulk (0.3-0.5 T) *(24,31)* and in bilayer $CrCl_3$ (0.23 T) *(27)*. A stable ferromagnetic ordering was not expected *a priori* when reducing the dimensionality of the few-layer AFM system down to the monolayer, as the removal of the interlayer exchange interaction does not guarantee that the intra-layer ferromagnetic behaviour remains intact. In fact, the magnitudes of the calculated magnetic anisotropy energies (MAE) for a single $CrCl_3$ monolayer were so low, that some calculations *(30,32)* even predict a marginally favorable perpendicular magnetic anisotropy (PMA) instead of the expected in-plane one. Our observations of an in-plane easy axis in monolayer $CrCl_3$ provide a clear answer to this puzzle and consolidates its classification as a two-dimensional in-plane ferromagnet. For completeness, we note that our few-layer $CrCl_3$ samples grown by MBE show similar magnetic behaviour as those reported in exfoliated flakes *(25-28)*, exhibiting signatures of spin-flop transitions at low fields *(28)* and giant interlayer exchange enhancement *(26,28)* in the form of large critical fields (Supplementary Figure 5).

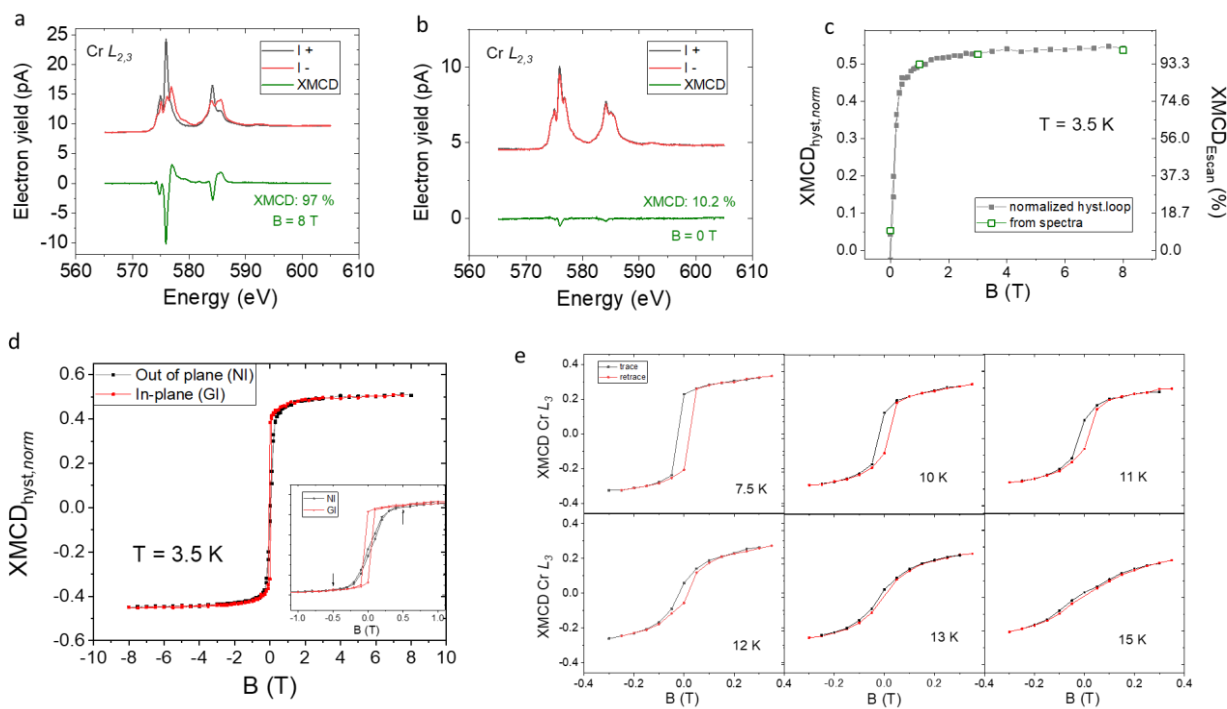

Figure 2. **Element-specific magnetic properties of a CrCl$_3$ monolayer measured by X-ray magnetic dichroism.** X-ray absorption spectra with different photon helicities and the resulting difference (XMCD), for the Cr L$_{2,3}$ edge at B = 8 T (a) and at zero magnetic field (b). A very large dichroic signal (close to 100%) can be observed at high fields. (c) Comparison between the XMCD magnitudes extracted from the energy spectra and from XMCD hysteresis loop sequences, showing consistency among both methods. (d) XMCD hysteresis loops taken in normal (NI) and grazing (GI) incidence, corresponding to the out-of plane and in-plane component of the magnetization, respectively. A snap of the low-field region is displayed as an inset to visualize the anisotropy fields (0.5 T). (e) XMCD signal as a function of in-plane magnetic field, taken at various temperatures. Remanence and coercive become negligible at 13 K, indicative of a magnetic phase transition.

To understand the origin of the unexpectedly large magnetic anisotropy, we analyse the relevant terms that apply to CrCl$_3$. Being a magnetic insulator with a local magnetic moment near 3 µB/Cr, the dominant magnetic interaction is a 90-degree ferromagnetic super-exchange hopping path over the Cl ligands, as expected from the Goodenough-Kanamori rules *(34,35)*. Considering this scenario, besides magnetocrystalline and shape anisotropy, a large effect of the ligand p-orbitals has recently been discussed *(36,37)*, in which the spin-orbit strength, the degree of covalency and the distortion angle from the octahedral configuration play an important role. The total anisotropy energy will be thus given by $\Delta E_K$ (total)= $\Delta E_{SI}$ + $\Delta E_D$ + $\Delta E_{ME}$, as defined by the magnetocrystalline (single-ion), dipolar and the magnetic exchange anisotropy terms, the latter related to ligand contributions in the Cr 3d- Cl p – Cr 3d super-exchange hopping path. We follow the rationale by Kim *et al. (37)* applied to CrI$_3$ to quantify the relevant energies. First, the single-ion anisotropy arises from the anisotropy of the orbital moment L, with an energy scale of $\Delta E_{SI}$ = - ξ $\Delta$L S/4, where ξ is the spin-orbit coupling strength. This term fully vanishes if the orbital moment is quenched, as expected in ionic-like Cr$^{3+}$ compounds. However, the Cr 3d spin-orbit (LS) coupling admixes the t$_{2g}$ and e$_g$ orbitals to recover a non-zero value of the orbital moment L. In fact, as extracted from sum-rule analysis (Supplementary Fig. 4), we obtain L=-0.08 µB for normal (out-of-plane) and L=-0.12 µB and grazing incidence (in-plane) XMCD spectra, respectively, yielding a measurable orbital anisotropy $\Delta$L= 0.04. Using ξ (Cr 3d) = 30 meV *(37)*, we obtain $\Delta E_{SI}$ = 0.14 meV in favour of an in-plane easy axis. As for the dipolar term, the in-plane easy axis is also favoured by the shape anisotropy and is of the order of $\Delta E_D$ =µ$_0$m$^2$/ 4πr$^3$ = 0.02 meV for Cr$^{3+}$ spins *(33)*. The last term ($\Delta E_{ME}$) derived in Ref. *(37)* and highlighted as a key factor to account for the large out-of-plane anisotropy in CrI$_3$, is strongly dependent on the ligand spin-orbit coupling strength $\xi_P$, the degree of covalency ($\Delta$) and the angular deviation $\delta\theta$ from the Cr-ligand plane given by the ideal crystal structure (degree of trigonal distortion), and follows a power-law scaling $\Delta E_{ME} \sim (\xi_P)^a (\delta\theta)^b / (\Delta)^c$. Compared to the parent compounds CrI$_3$ and CrBr$_3$, it is evident that CrCl$_3$ attains the lowest magnetic exchange anisotropy due to the low spin-orbit strength of the Cl ligand and smallest degree of covalency (largest bandgap). Considering values of $\xi_P$=70meV, $\Delta$=3.5eV and $\delta\theta$=2° and using four-site multiplet cluster calculations, the calculated $\Delta E_{ME}$ in *(37)* resulted in negligible values (1 µeV) for CrCl$_3$. In our case, based on the experimental STS data (Fig. 1d) the bandgap of CrCl$_3$ is estimated to be 1.6 eV, resulting in a factor of 0.45 in the charge transfer energy $\Delta$ and a factor of 4 (using the exponent c =1.73) increase of $\Delta E_{ME}$ with respect to the calculations, being still of the order of a few µeV. The main message here is that only a substantial trigonal distortion ($\delta\theta$) will result in a sizable $\Delta E_{ME}$ for our MBE-grown CrCl$_3$ layers. In fact, we observe by angular-dependent X-ray absorption measurements (Supplementary Figure 6) a maximum in the pre-edge intensity of the Cr L$_3$ edge at $\theta$=25° degrees instead of 35.3° to the normal plane, the latter expected from the ideal octahedral structure. This measurement is sensitive to the admixed Cr 3d-or6itals (e.g. d$_{3x2-r2}$) within the Cr-Cl bonding plane, and is therefore a good tool to map an angular deviation. We believe that this substantial trigonal distortion ($\delta\theta$ = 10°) could be induced in our samples during the nucleation and growth process, and increase not only $\Delta E_{ME}$ by an order of magnitude (~40 µeV) but also impact the anisotropy of the orbital moment ($\Delta L$=0.04) that enters the single-ion anisotropy term ($\Delta E_{SI}$) and which is the dominant contribution to explain the observed anisotropy fields in the XMCD hysteresis loops.

In order to characterize the magnetic behaviour below and above T$_c$ and find insights into the nature of the phase transition, the low-field dependence of the Cr L$_3$ edge XMCD was acquired between 3.5 K and 15 K in the easy axis direction (for the detailed datasets see Supp. Figure 7). Concomitant with the disappearance of the hysteresis at 13 K (Figure 2e), the M(H) curves evolve into a softer S-shape above T$_c$ and move towards a linear dependence, entering the paramagnetic regime. As for the study of the critical behaviour at the phase transition, the magnetization is a good order parameter and scales as M= M$_0$ (1-T/T$_c$)$^\beta$, $\beta$ being the critical exponent that determines the universality class. In two-dimensional systems, the isotropic Heisenberg system (n=3) does not have any spontaneous order, such that the Ising

(n=1) and XY (n=2) models are the relevant scenarios for $CrCl_3$, where n is the dimensionality of the spin degree of freedom. Though often understood in terms of an out-of-plane easy axis, the Ising universality can also be found in in-plane magnetized systems *(38,39)*, as long as there is only one preferred (uniaxial) magnetization direction. To perform the analysis of the critical exponent β, we rely on the evolution of the remanent magnetization $M_r$ (XMCD signal at zero field) as a function of temperature, as shown in Figure 3a. The inset shows the fitting of the data close to the phase transition (12-13K) yielding values of (β=0.227± 0.021) and $T_c$ = (12.95±0.03) K, matching well with the expected value (β=0.231) of the 2DXY model *(9)*. If the exponent β is set to equal 0.125 corresponding to Ising-type universality (green line in Figure 3b), there is very poor agreement with the experimental data. A better visualization of the extent of the temperature region where a certain critical behaviour holds is found through a log-log representation of the magnetization versus reduced temperature (1-$T/T_c$), as shown in Figure 3c. The region where the critical scaling holds occurs in a range of reduced temperatures between 0.1 and 0.003, a regime typically used in critical phenomena studies (for a compendium see Ref. 14). To further assess the 2D-XY behaviour in our samples, a second analysis procedure, the so-called Arrott-Noakes plots *(40)*, has been carried out from the field- and temperature dependent XMCD data. In this representation, the magnetization and susceptibility power-law scaling are visualized together across the phase transition, and a consistent set of the critical exponents β and γ, corresponding to the magnetization and susceptibility, can be deduced. This approach is widely used to distinguish between the various magnetic interaction models (e.g. Heisenberg, Ising, XY, Potts) in 3- and 2-dimensions, each of which have a specific set of critical exponents. On this basis, we pursue the determination of the critical exponent γ by an Arrott-plot analysis of the temperature dependent XMCD data, as shown in Figure 3d. The analysis of the Arrott-plots is consistent with the previously inferred value of β (0.227) and in addition, define the susceptibility critical exponent as γ=2.2, the latter falling reasonably in the range estimated from the high-temperature series expansions for the 2DXY model γ=2.4 ± 0.3 *(41)*. It should be noted, however, that although the linear fits and the evolution of the intercepts are reasonable for these set of critical exponents, the Arrott-plot analysis is not as accurate as the temperature dependent magnetization to determine the Curie-Temperature $T_c$, i.e. the high field data of 12.5 K and 13 K fall under the same fitting range. (More details about issues in determining the susceptibility exponent γ of 2DXY systems can be found in the Supplementary Information). It is important to underline that these results have been reproduced on a second sample with a slightly different $T_c$ but with the same XY universality scaling, demonstrating that monolayer $CrCl_3$ grown by MBE is a robust easy-plane magnet (see Table 1 and Supp.Figure 8 for the additional datasets). The very good agreement of the critical exponent β (and the qualitative one of γ) to the theoretically expected value, is attributed to the ideal 2DXY features of our samples: vanishing in-plane crystalline anisotropy fields, negligible substrate interaction and a relatively large magnetic coherence size L of the $CrCl_3$ monolayer due to a homogeneous, atomically flat morphology on the graphene substrate. (Note that all other 2D universality scaling models fail to describe our data, see Supplementary Figure 9). On a more general footing, the observed 2DXY ordering in monolayer $CrCl_3$ suggests that the nature of the phase transition is of Berezinskii-Kosterlitz-Thouless (BKT) type. This universal phase transition has been observed in other physical systems such as superconductors *(42,43)*, atomic surface reconstructions *(44)* and quasi-condensates of atomic gases *(45)*. In magnetic systems, this transition has been argued to occur in metallic monolayers grown on crystalline substrates *(46,47)* and in layered bulk magnets *(48)*. However, the strong substrate interaction in the first, and interlayer exchange in the second case, might cause a departure from ideal 2DXY behaviour where the BKT transition is expected. We believe that our $CrCl_3$ system, a quasi-free-standing van der Waals monolayer, is a more suitable playground for a magnetic BKT transition to take place. Further studies on finite-size effects and exponential scaling of the magnetic correlation length are needed to fully assess this conjecture. In a very recent theoretical work, a transport signature of the BKT transition in magnetic easy-plane systems has been formulated *(49)*, which perfectly applies to our $CrCl_3$ monolayer and shall bring additional insights on the spin transport phenomena of 2D easy-plane magnets, expected to reach the spin superfluidity regime *(50)*.

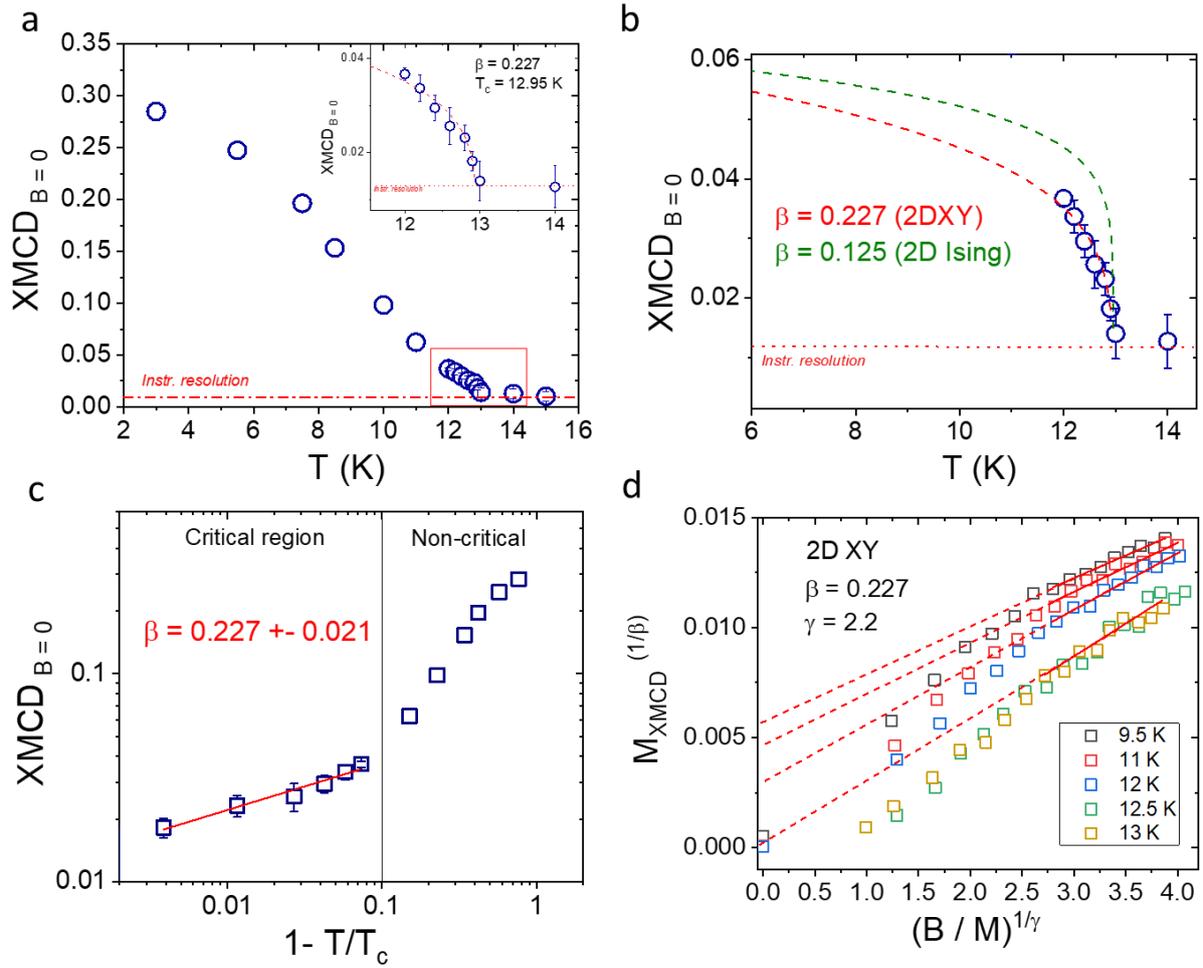

Figure 3. **Scaling behaviour and critical exponents.** (a) XMCD values at zero magnetic field (remanence) as a function of temperature, extracted from the hysteresis loops. The critical scaling fit close to the phase transition is shown in the inset, determining the exponent β=0.227. (b) The fits using the 2D Ising (green) and 2DXY models (red) are shown, being the latter the one consistent with our data. (c) Logarithmic representation of XMCD magnitude vs reduced temperature (1-T/$T_c$), for an optimal visualization of the accuracy of the inferred critical exponent β and the critical temperature region. (d) Modified Arrott-Plots of the field dependent XMCD data at various temperatures (β=0.227, γ= 2.2), matching with the 2DXY model predictions. Extrapolation of the linear fits at high-fields is drawn as guide to the eye to visualize the y-intercept evolution towards the phase transition (y=0 at $T_c$).

Taking into account the analysis presented above, we have demonstrated that a $CrCl_3$ monolayer, featuring van der Waals coupling to the graphene substrate and hexagonal symmetry in the plane but without strict single crystalline long-range order, constitutes a nearly ideal realization of a 2DXY magnetic system. The van der Waals nature at the $CrCl_3$/substrate interface minimizes effects such as hybridization, bonding, and substrate-driven crystalline anisotropy fields, factors that can drive the system to an Ising or anisotropic Heisenberg system; or even cause a departure from a 2-dimensional behaviour by strong substrate hybridization. The in-plane hexagonal crystal symmetry (XYh6) of $CrCl_3$, on the other hand, is the least perturbative for the XY model *(6)* and is additionally diminished in our experimental system due to the in-plane twisting of crystalline domains. As compared with bulk layered magnets, our monolayer system has no interplanar exchange coupling $J_{ex,inter}$ which is also a perturbation of ideal 2DXY behaviour and dimensionality. Hence, our quasi-freestanding, atomically-thin $CrCl_3$ layer grown by MBE, constitutes an appealing model system that can, in turn, be treated accurately by theoretical calculations. Interesting follow-up studies would be to gently turn on in-plane anisotropy fields by choosing different rigid substrates with strong crystalline fields or a given step terrace morphology, and see how this affects the 2DXY behaviour of the monolayer. In that way, recipes to attain a crossover to Ising-type (uniaxial anisotropy) behaviour may be developed, a useful pathway to design on-demand magnetic anisotropies for functional spintronic devices based on 2D materials. On a more fundamental side, the finite-size effects of the BKT phase transition can be studied by modifying the grain size and percolation behaviour of the MBE-grown monolayers. Moreover, an epitaxy-controlled trigonal distortion could be used to set the balance between single-ion anisotropy and dipolar interaction in $CrCl_3$, crucial to stabilize topological spin textures -such as merons- *(51,52)* in two-dimensions. These tuning knobs are a powerful tool to reach beyond the state-of-the-art and current understanding of van der Waals magnets prepared by exfoliation. Our demonstration of a $CrCl_3$ ferromagnetic monolayer grown by MBE is thus a critical step to achieving exquisite control of the magnetic properties and spin textures of innate 2D systems by thin film growth engineering.

**Acknowledgements**

A.B-P. thanks the HZB and CELLS-ALBA for the allocation of synchrotron radiation beamtime under proposals 192-08773-ST (HZB) and 2019093862 (CELLS-ALBA). We warmly thank Felix Küster for assistance on the LT-STM measurements and Chen Luo, Kai Chen, Sangeeta Thakur and Steffen Rudorff for technical support at BESSY.

**Funding**

F.R and A.B-P. acknowledge the financial support for the VEKMAG project and for the PM2-VEKMAG beamline by the German Federal Ministry for Education and Research (BMBF 05K10PC2, 05K10WR1, 05K10KE1) and by HZB. MV and PG acknowledge additional beamtime through the ALBA IHR program and funding via Mineco grant FIS2016-78591- C3-2-R (AEI/FEDER, UE). K.C. was funded by National Natural Science Foundation of China (Grant No. 12074038).

**Author contributions**

A.B-P., K. C. and S.S.P.P. conceived the study, and A.B-P. was the lead researcher. A.B-P. carried out the full magnetic characterization by XAS/XMCD, analysed the data and wrote the manuscript. J.J. grew the samples for the beamtimes under the guidance of A.B-P. and K.C, and performed the in-situ RHEED and STM characterization. K.C. initiated and optimized the substrate preparation and epitaxial growth. P.S. performed the low-temperature STM characterization. J.J., A.K.P, J.T. and F.R. assisted with the XMCD measurements in BESSY, whereas P.G. and M.V. assisted and performed part of the XAS/XMCD experiments in ALBA. All authors discussed the data and commented on the manuscript. S.S.P.P. supervised the entire project.

## Supplementary Materials

Materials and Methods

Figures S1 to S9

Supplementary Note 1